\documentclass[lettersize,journal]{IEEEtran}
\usepackage{amsmath,amsfonts}
\usepackage{algorithm}
\usepackage{array}
\usepackage[caption=false,font=normalsize,labelfont=sf,textfont=sf]{subfig}
\usepackage{textcomp}
\usepackage{stfloats}
\usepackage{url}
\usepackage{verbatim}
\usepackage{graphicx}
\usepackage{cite}
\usepackage{dialogue}
\usepackage{tikz}
\usepackage{algpseudocode}
\usepackage{hyperref}
\usepackage{booktabs}
\newcommand\encircle[1]{%
\tikz[baseline=(X.base)] 
   \node (X) [draw, shape=circle, inner sep=0pt, text depth=1pt, fill=black, text=white,] {\strut \small{#1}};}
\newcommand{\systemname}[0]{One For All}

\MakeRobust{\Call}

\begin{document}

\title{One Agent Too Many: User Perspectives on Approaches to Multi-agent Conversational AI}

\author{
  Christopher Clarke$^1$\hspace{10pt} Karthik Krishnamurthy$^2$\hspace{10pt} Walter Talamonti$^2$ \\ \hspace{10pt} Yiping Kang$^1$\hspace{10pt} Lingjia Tang$^1$\hspace{10pt} Jason Mars$^1$   \vspace{0.3cm}\\
    \text{$^1$University of Michigan, Ann Arbor, MI}\\
    \text{$^2$Ford Motor Company, Dearborn, MI}\\
    \text{\{csclarke, ypkang, lingjia, profmars\}@umich.edu} \\
    \text{\{kkrish65, wtalamo1\}@ford.com}
}


\markboth{Journal of \LaTeX\ Class Files,~Vol.~14, No.~8, August~2021}%
{Shell \MakeLowercase{\textit{et al.}}: A Sample Article Using IEEEtran.cls for IEEE Journals}


\maketitle

\begin{abstract}
Conversational agents have been gaining increasing popularity in recent years. Influenced by the widespread adoption of task-oriented agents such as Apple Siri and Amazon Alexa, these agents are being deployed into various applications to enhance user experience. Although these agents promote \textit{``ask me anything''} functionality, they are typically built to focus on a single or finite set of expertise. Given that complex tasks often require more than one expertise, this results in the users needing to learn and adopt multiple agents. One approach to alleviate this is to abstract the orchestration of agents in the background. However, this removes the option of choice and flexibility, potentially harming the ability to complete tasks. In this paper, we explore these different interaction experiences (one agent for all) vs (user choice of agents) for conversational AI. We design prototypes for each, systematically evaluating their ability to facilitate task completion. Through a series of conducted user studies, we show that users have a significant preference for abstracting agent orchestration in both system usability and system performance. Additionally, we demonstrate that this mode of interaction is able to provide quality responses that are rated within 1\% of human-selected answers.

\end{abstract}

\section{Introduction}
The conversational AI market is growing at an increasingly rapid pace and is projected to reach a valuation of US \$13.9 billion by 2025~\cite{market_and_markets_2020}.
The growth of this market, largely influenced by the popularity of task-oriented agents such as Apple Siri, Google Assistant, and Amazon Alexa, has led to an increased interest in the use of these conversational agents in domain-specific areas such as driver assistance~\cite{Lin2018Adasa}, home automation~\cite{Luria2017}, and food ordering~\cite{foodbot}.
Companies are rapidly building conversational agents dedicated to their own product offerings, providing an entirely new form of interaction for their customers.
Today, platforms such as Whatsapp and Facebook host more than 300,000 of these conversational agents ~\cite{Chaves:2018:SMC:3173574.3173765,nealon_2018}.

The majority of the commercially available agents today are designed to perform tasks in highly specialized domains \cite{roller2020recipes, clarke-etal-2022-one}.
For example, a user may use an agent like Amazon Alexa for online shopping but engage with Google Assistant for daily news updates. Additionally, a given agent may be more proficient at a specific domain over another e.g A finance CA is better suited to answer finance questions. As a result, when navigating complex tasks, users are tasked with learning and adopting multiple agents \cite{clarke-etal-2022-one}.

An example of this multi-agent interaction mode in practice is demonstrated by the Amazon Voice Interoperability Initiative\footnote{\url{https://developer.amazon.com/en-US/alexa/voice-interoperability}}. This agreement between agent vendors provides users with choice and flexibility through the utilization of multiple intelligent assistants on voice-enabled products such as smart speakers and smart displays \cite{amazon.com}. Despite its potential benefits, prior work has presented challenges with the cognitive burden of comprehending each agent's capability and navigating between agents. This coupled with the continuous growth of conversational agents leads to decision overload for users\cite{dubiel2020, novick2018, choice2019, clarke-etal-2022-one}. However, other studies suggest this to be a non-issue and in some cases show enhancement of user experience \cite{cui-guideline, amazon.com, Chaves:2018:SMC:3173574.3173765}. 


Despite the employment of both of these interaction modes in practice, to the best of our knowledge, no prior work has explicitly investigated the impact of these multi-agent interaction modes on user decision-making and cognitive load. In addition, a study is also needed to identify the key constraints and challenges of designing and employing such interactions. Thus, it remains unclear which interaction mode is best suited to support user interaction and provide the best overall performance.

Bridging the interaction gap of using multiple conversational agents spanning different domains presents several key challenges.

\begin{itemize}
    \item First is the assurance that the user is routed to the best available agent for their particular query. Expert users of multiple conversational agents over time build a mental model of each agent and its capabilities to facilitate their decision-making. Single-agent interfaces in contrast face the challenge of understanding which agent is best suited for a given task.
    \vspace{0.2cm}
    \item Second, these conversational agents are constantly improved upon and expanded with new capabilities. Any orchestration approach needs to be flexible and adaptive to these changes with relative ease. For individual users, this adaption is built through use over time. In contrast, any agent integration approach, given that these agents are black boxes, must operate without relying on the internals of any given agent whilst having a general sense of each agent's capability and being adaptive to its change.
    \vspace{0.2cm}
    \item Lastly, is the ability to cover the entire spectrum of domains supported by all of the agents in the ensemble.
\end{itemize}

In this paper, we investigate the interactional experience of multi-agent conversational AI via single-agent use (one agent for all) vs multi-agent use (user choice of agents) in terms of usability and overall system performance. For this purpose, we implement two conversational agent prototypes to simulate each interaction experience which denote as \systemname{} and Agent Select respectively. To evaluate both experiences we conducted a series of user studies consisting of 19 participants, in which participants completed a series of tasks across several domains using both prototypes.

Our main contributions are as follows:

\begin{itemize}
    \item An analysis of the interaction experience of multi-agent conversational AI via single agent use (one agent for all) vs multi-agent use (user choice of agents) in terms of usability and system performance. We show that users have a significant preference for the use of a single-agent interface over multi-agent in regards to system usability and system performance.
    \vspace{0.2cm}
    
    \item We release our end-to-end implementation of prototypes One For All and Agent Select for community use in exploring conversational user interfaces in the context of multi-agent support. Additionally, we release two datasets. The first contains questions asked by our study participants, the returned agent response, and the subjective user satisfaction (yes/no) on the acceptability of the agent response available. The second consists of utterance response pairs collected from our system evaluation accompanied by our curated crowdsourced human judgments \footnote{\url{https://gitlab.com/ChrisIsKing/one-for-all}}.
    \vspace{0.2cm}

    \item We explore key design and system challenges of employing each interaction mode and provide a series of insights and suggestions for optimizing the interactional experience.
\end{itemize}

\section{Related Work}


Researchers have long explored the realization of interacting with machines in a conversational manner.
Such exploits can be traced as far back as the 1960s, with applications such as Eliza and the discussion of Human-Computer Symbiosis~\cite{Weizenbaum:1966:ECP:365153.365168, Licklider:1992:MS:612400.612433}. Modern conversational systems have made significant progress in recent years~\cite{Fast_2018, Cassell:2000:ECI:332051.332075, Subramaniam:2018:CCM:3237383.3237472, Lin2018Adasa, Porcheron:2018:VIE:3173574.3174214}, bridging the human-machine interaction gap, particularly through voice user interfaces~\cite{Porcheron:2018:VIE:3173574.3174214}. However, despite this exponential growth in CAs, their range of support is still limited to a single or set of domain areas. In this paper we explore modern approaches employed in practice to extend this range of support. We build upon prior work in NLP and HCI  to extend upon the capabilities of  current conversational agents by leveraging multiple existing agents from different domains. This section discusses two main related areas of research: (1) Multi-agent Conversational systems and (2) The use of conversational AI in HCI.

\textbf{Multi-agent systems:} The use multiple agents in conversational AI is a fairly unexplored area. Chaves and Aurelio~\cite{Chaves:2018:SMC:3173574.3173765} compared the interactional coherence in the scenarios of a multi-party agent conversation and a single agent conversation. By simulating both single and multi-agent settings in a chatbot environment they found that users reported more confusion in multi-chatbot interactions. Indicating that even in a text-based environment, information is best consumed through a single source. Subramaniam et al.~\cite{Subramaniam:2018:CCM:3237383.3237472} describe in their work a conversational framework that employs what they call an \textit{Orchestrator Bot} to understand the domain user query and provide a response either from a domain expert or its own knowledge base. This domain prediction is achieved by applying linguistic analysis of the query to extract information based on each of the supported domains. Additionally, systems such as Alana \cite{alana} explore this ensemble integration using a combination of rule-based heuristics for priority based bot selection. However, in practice, these CAs are "black boxes" whose inner-workings and supported domains are not public or explicit. Thus limiting the application of such an approach in practice. Lasecki et at. \cite{lasecki2013chorus} in \emph{CHORUS} explore multi-agent integration as a human crowd-powered tool by leveraging multiple crowd workers to propose and vote on responses. 

Services such as Alexa Skills for Amazon Alexa and Google Actions for Google Assistant have made attempts to provide developers with toolkits designed to extend the functionality of their agent offerings and allow multi-agent use. However, these services pose three problems: (1) developers are constrained to using provided toolkits, (2) agent deployment is limited to a subset of physical devices, and (3) interaction with extended features requires invocation by the user (e.g., \emph{``Alexa, launch flight info app.''}. Clarke et al. \cite{clarke-etal-2022-one} in their work on multi-agent conversational AI address these concerns by developing an agent and device agnostic approach that requires no additional end-user expertise to leverage and explore the approaches of question-agent pairing and question response pairing for multi-agent integration. In our work, we employ a question-response pairing approach to designing \systemname{} as recommended by \cite{clarke-etal-2022-one}.

\textbf{Conversational AI in HCI:} Recent studies in HCI~\cite{Cassell:2000:ECI:332051.332075, chang2019, scuito2018, heloisa2020} highlight the need to further examine the use of conversational agents in everyday life. Methods such as interviews, ethnomethodology, usability, and longitudinal studies have all been used to evaluate conversational agents~\cite{Reeves:2018:VCU:3170427.3170619, hasan:2018}. Other methods such as in-home deployments, conversation analysis, and deep learning are also being put into practice~\cite{jaganath:2018,huang2016there}. Scuito et al \cite{scuito2018} in their work explore the in-home conversational agent usage of Amazon Alexa showing just how ubiquitous CAs have become revealing that most users interact with Alexa on a daily basis. However, despite this widespread adoption the range of support still remains fairly limited and users spent a lot of time trying to get the expected response from conversational agents \cite{lee21, luger16}. This further highlights the need for multi-agent systems capable of expanding the range of supported tasks. We showcase the applicability of this by using commercially available CAs and comparing the user experience to the status quo of using multiple agents.

\begin{figure*}
  \centering
  \includegraphics[width=1\columnwidth]{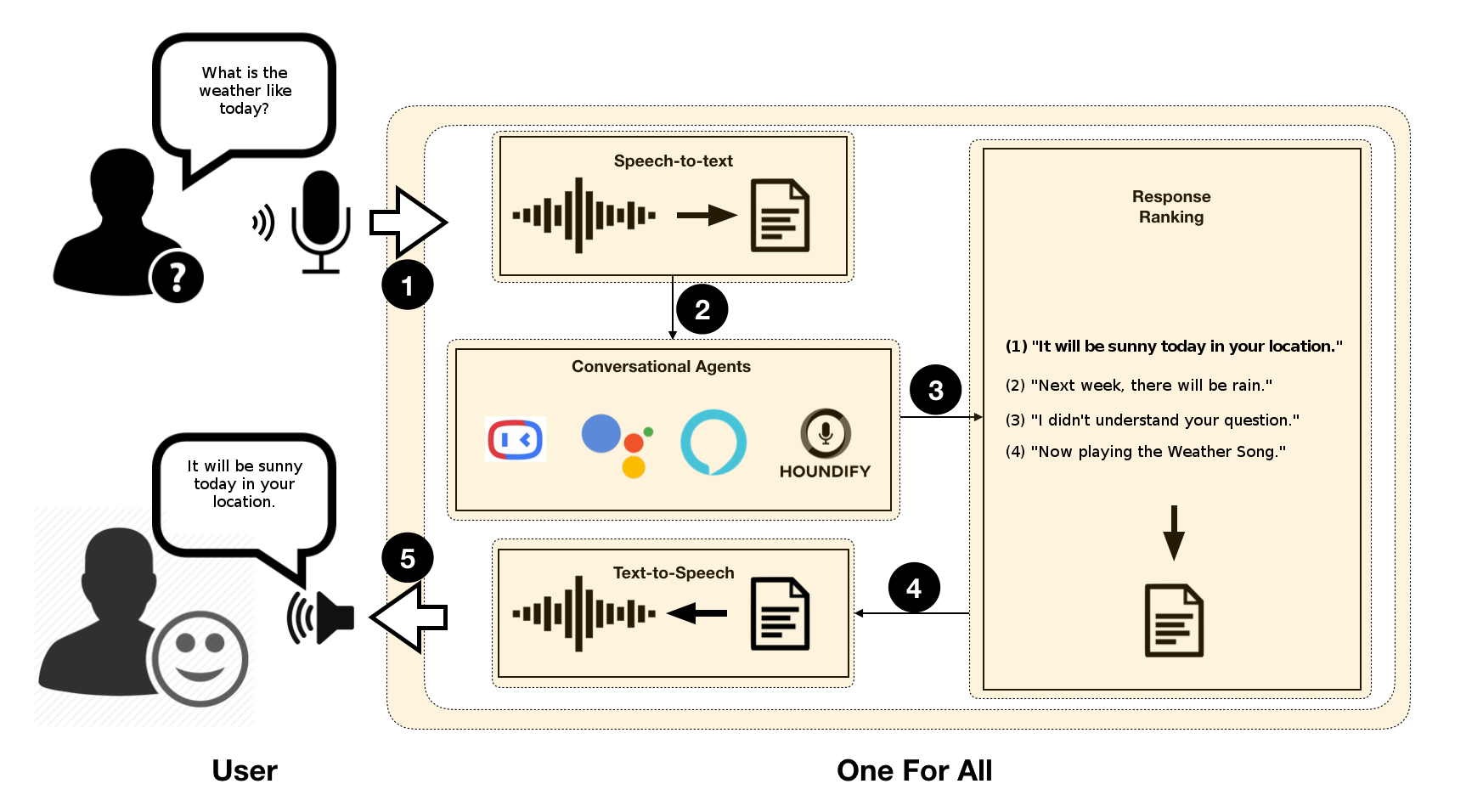}

  \caption{Overview of \systemname{}.
The user's voice is captured via a microphone and then transcribed into its text equivalent (\protect\encircle{1}).
The textual query is then passed to each of the conversational assistants in parallel (\protect\encircle{2}).
When each agent provides a response, all responses and the input query are fed to the ranking engine (\protect\encircle{3}), which then embeds them and calculates a \textit{semantic relation} between the user's query and agent responses.
The response with the highest ranking is returned (\protect\encircle{4}), then converted to audio and played to the user (\protect\encircle{5}).}~\label{fig:SystemOverview} 
\end{figure*}

\section{System Design}
In this section, we detail the overall architecture of our two prototypes: \textit{\systemname{}} and \textit{Agent Select}. We present in detail each of the major software components and illustrate how a query is processed in the system workflow. 

\subsection{Conversational Agents}
To explore the interactional experience of multi-agent use, we integrate a suite of commercially available agents consisting of Amazon Alexa\footnote{\url{https://developer.amazon.com/en-US/alexa}}, Google Assistant\footnote{\url{https://assistant.google.com/}}, SoundHound Houndify\footnote{\url{https://www.houndify.com/}}, and Ford Adasa~\cite{Lin2018Adasa}. These agents were selected due to their wide range of supported domains, popularity in the user market, and their provision of open programming interfaces to allow their integration. More information on each of the respective agents is presented in appendix \ref{agents}.



    


\subsection{\systemname{}}
\systemname{} provides a voice-enabled interface for users to ask questions to the ensemble of conversational agents. To facilitate this integration \systemname{} provides each agent in the ensemble the opportunity to respond to the query and returns the best-ranked response to the end user. In order to facilitate this response ranking, we consider three state-of-the-art models in NLP models to generate contextual vector representations (called \emph{embeddings}) for the user query and agent responses. These models were chosen for their proficiency in generating feature-rich textual representations for a variety of NLP downstream tasks. We use these vector representations to calculate the \textit{semantic relation} between the query and candidate responses and return the top-ranked response to the user.

Figure~\ref{fig:SystemOverview} presents a high-level overview of the system components and how queries are handled in \systemname{}.
\systemname{} is comprised of several software component runtimes executing in parallel to achieve a unified experience. The following subsections describe each component in detail.

\vspace{0.1cm}
\subsubsection{Speech to Text Engine}
The first major component of \systemname{} is the speech-to-text interface.
This interface allows users to ask questions vocally in natural language, providing a seamless way for users to interact with \systemname{} and its supported agents.
\systemname{} uses Google Speech API \cite{tackstrom2013token} for Automatic Speech Recognition, which is widely deployed in production environments \cite{Hauswald:2015:SOE:2775054.2694347, hasan:2018}.
As described in our evaluation, we observed a 0\% transcription error using this service during our experimentation, consistent with performance observed in production~\cite{googleASRnumber}.
This speech-to-text API first processes and extracts feature vectors representing the voice segments and then submits these feature vectors to a speech recognition kernel to transcribe users' utterances.
The transcribed text then serves as the input to the next component. Each of the conversational agents receives the user's input query as text and then provides a response to that query. Each response is collected and passed to the response ranking engine.


\vspace{0.1cm}
\subsubsection{Response Ranking Engine} 
Once all candidate responses are collected, we rank each agent response by its semantic relation to the original input query. For example, if a user asks about the weather, the system should highly rank weather-related responses and ignore responses that relate to other domains (e.g., a car-related response would be inappropriate for a weather-related inquiry).
The NLP models considered are as follows:

\begin{itemize}
    \item BERT~\cite{devlin-etal-2019-bert}: a multi-layer Bidirectional Transformer model that learns universal representations of text and obtains state-of-the-art results on an array of language processing tasks. We used the Bert as a Service\footnote{\url{https://github.com/hanxiao/bert-as-service}} encoding service using embeddings from the second to last encoding layer.
    \vspace{0.1cm}
    
    \item Universal Sentence Encoder~\cite{cer-etal-2018-universal}: A sentence encoding model for encoding sentences into high dimensional vectors that can be used for text classification, semantic similarity, clustering and other natural language tasks. We use the transformer model\footnote{\url{https://tfhub.dev/google/universal-sentence-encoder-large/3}} for our system.

    \item Smooth Inverse Frequency (SIF)~\cite{arora2017asimple}: A weighted average of word embeddings, with weights determined by word frequency within a corpus. We used pretrained, 50-dimensional Glove~\cite{pennington2014glove} embeddings for our implementation. Glove is an unsupervised learning algorithm for generating word embeddings by aggregating global word-word co-occurrence matrices from a corpus.  
\end{itemize}

\begin{algorithm}
\caption{Response Ranking}\label{alg:rank}
\begin{algorithmic}
\Require{$R_{1} \dots R_{N}$, $Q$} 
\Ensure{$r_{i}$ (highest ranked response)}
\Function{rank}{$Q$, $R$}:
\State {$A$ $\gets$ {$[]$}}
\State {$N$ $\gets$ {$length(R)$}}
\For{$i \gets 1$ to $N$}                    
        \State {$A[i]$ $\gets$ {\Call{Dist}{\Call{Embed}{$Q$}, \Call{Embed}{$R[i]$}} }} \Comment{Calculate distance between vector representations}
\EndFor
\State \Return {$R[\Call{argmin}{A}]$}
\EndFunction
\end{algorithmic}
\end{algorithm}

Our response ranking algorithm is as follows:

\begin{enumerate}
    
    \item Generate a vector representation (i.e., an embedding) of the user's query and each of the $1 \dotso n$ agent responses.

    \item For each of the $n$ agent responses, we calculate the Euclidean distance between the agent response and the query.
    
    \item Rank by distance in ascending order.
\end{enumerate}

The intuition behind this approach is to capture the semantic relation between the user's query and the agent's response. This acts as an indicator of how well the agent was able to understand the user's query.
A response that produces poor semantic relation indicates that the agent was either unable to understand the query or responded poorly to the question.
For example, consider the input query \emph{``What is the weather outside?''} and a candidate response \emph{``Sorry, I don't know how to help with that.''}.
When embedded, these two sentences will have a very low relation score.
In contrast, an alternate candidate response \emph{``The weather outside is delightful.''} will appear closer in the embedding space to the original query, and is then ranked higher.
Additionally, utilizing a response selection approach, allows us to address the aforementioned challenges of adapting to agent change over time. Since this method requires no training and does not rely on the internals of the agents themselves, it is applicable even if we add or remove agents from the ensemble.

\subsubsection{Text-to-Speech Engine} Once the highest ranked response is selected, it is then converted to audio to be played back to the end user. \systemname{} uses the Google Cloud Text-to-Speech Synthesis Library~\cite{tackstrom2013token}. Voice attributes such as gender, pitch, volume, and speech rate are configurable via the service\footnote{We use the standard Google Female English voice with all other parameters set to default.}.

\subsection{\systemname{}: Life of query}
To illustrate how these components are integrated, we walk through the lifetime of a query step-by-step following the circled numbers in Figure~\ref{fig:SystemOverview}.
The life of a query begins with a user's voice input---the user activates the system to begin listening via a key press. The captured voice input is sent to the Speech-to-Text engine of \systemname{} (\encircle{1}) and then transcribed to its text equivalent.
The query data\footnote{Depending on each agent's input requirements, the data is sent as a transcription or as raw audio data.} is sent to each conversational agent in parallel (\encircle{2}). The system awaits a response from each of the CAs before passing the responses to the response ranking engine.
The user's query and the agents' responses are passed to the ranking engine (\encircle{3}), where they are converted to vector representations (embeddings), and the \textit{semantic relation} is calculated as a basis of ranking the responses.
The best-ranked response is then returned to a Text-to-Speech engine to convert it to audio that is played back to the user in response to their query.

\begin{figure}
  \centering
  \includegraphics[width=1\columnwidth]{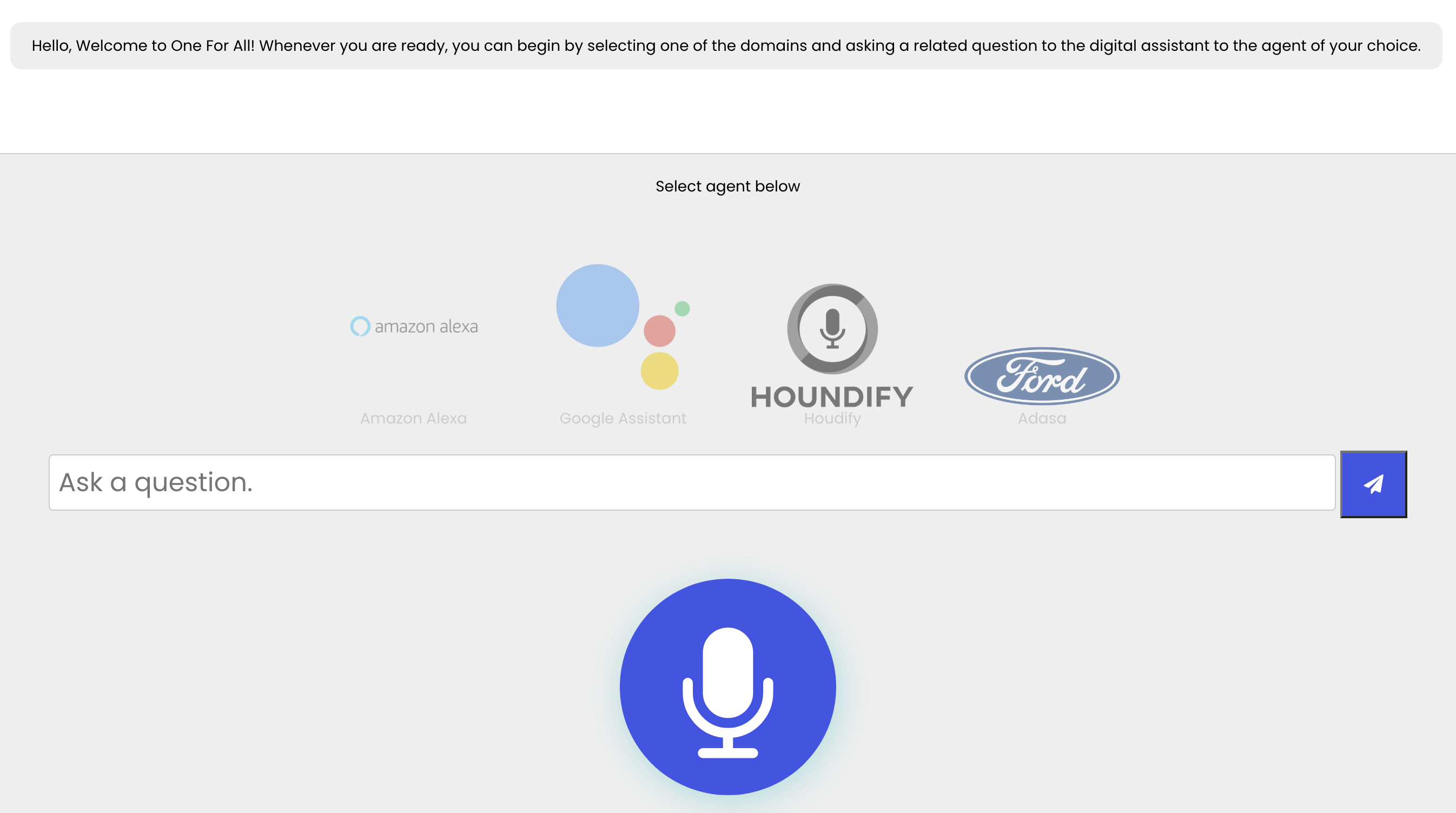}
  \caption{Agent Select Prototype Interface. In contrast to \systemname{}, users are given a choice over the agent they wish to route their query to.
  }~\label{fig:agent-select}
\end{figure}

\subsection{Agent Select: Life of query}
As shown in figure \ref{fig:agent-select} Agent Select provides a voice-enabled interface for users to ask questions to any of the integrated conversational agents via a computer interface. It re-uses the same components as \systemname{} without the need for a ranking engine since the user selects their desired agent beforehand. The life of a query begins with the selection of a desired agent to query. The user first clicks the agent of their choice. Second, the user activates the system to begin listening via a key press. The captured voice input is sent to the Speech-to-Text engine of \systemname{} (\encircle{1}) and then transcribed to its text equivalent. The query data is sent to the CA of their choice and the agent response from the chosen CA is returned to the user.

 \begin{table*}[t]
\centering
\small
\caption{Targeted assessments of questions in questionnaires for each experiment.}
\label{tab:questable}
\begin{tabular}{c p{6cm} p{6cm}}
\toprule
\textbf{Statement} & \textbf{Agent Select} & \textbf{\systemname{}}\\
\midrule
 S1 & Being able to explicitly select which assistant is useful. & Having the system understand which assistant is responsible is useful.\\
 S2 & The answers provided by the digital assistant were useful. & The answers provided by the digital assistant were useful. \\
 S3 & The assistant is capable of understanding my question and providing a feasible answer. & The assistant is capable of understanding my question and providing a feasible answer. \\
 S4 & Having multiple assistants make the experience more pleasurable. & Having one unified assistant makes the experience more pleasurable. \\
 S5 & I am very satisfied with my experience using the prototype. & I am very satisfied with my experience using the prototype. \\
\bottomrule
\end{tabular}
\end{table*}

\section{User Study}
To gain insight into user perceptions of the different interaction modes we conducted a pilot study consisting of 19 participants. We issued a general call for participants in the Computer Science department with the aim of attracting a base of participants most likely to possess relevant conversational agent usage experience. 58\% of the participants reported using their digital assistant 1--5 times per day while 32\% indicated no use of any digital assistant as a part of their routine daily activity. Thus showing a fair distribution of regular CA users and non-users. Of the 4 integrated agents for this experiment most expressed familiarity with Amazon Alexa and Google Assistant. No participants were familiar with Adasa and only one participant had knowledge of Houndify.

\subsection{Experiment Prototype}

We deployed both prototypes on a computer system for users to interact with each interface to complete their tasks. Our experiment was built to compare two classes of usages: (1) single agent interface using \systemname{} as described in figure \ref{fig:SystemOverview}, and (2) multi-agent interface in which participants choose which agent they wanted to interact with as shown in figure \ref{fig:agent-select}.

\subsection{User Study Protocol}
Upon participant's arrival to the research lab, each participant was greeted and briefed that they would be participating in a study investigating the interactional experience of multi-agent conversational systems. They then proceeded to fill out a pre-screening questionnaire which recorded basic demographic information and consisted of questions centered around their previous experience and usage of digital assistants. Upon completion of the screening, participants were given an overview of each of the integrated conversational agents in the prototypes and presented with a sample task list as a reference. After this, the researchers explained the task instructions.
Time was provided for participants to familiarize themselves with each of the agents and the task instructions in addition to asking any questions. Once participants indicated their readiness and understanding of the task at hand, researchers commenced the experiment. Upon completion of each experiment, participants were asked to fill out a questionnaire regarding their experience using each prototype and complete a Software Usability Scale (SUS) test. When participants confirmed they had no further questions or feedback, the study was complete.

\subsection{Tasks}
Users were required to evaluate both prototypes \systemname{} and Agent Select. Before conducting the study, participants were briefed on the capabilities of the agents available to them as well as the mode of interaction for each of the prototypes. Each user was tasked with interacting with both prototypes executing a series of tasks spread across 10 domains. In each phase, users were instructed to complete a \textbf{minimum of at least 10 tasks}. Each task entailed interacting with the prototype and reporting if the task was completed as expected. The outline of tasks done for each experiment is depicted in the following subsections.

\subsubsection{Task list}
Each participant was presented with a sample task list comprising of 50 example queries in the domains of Automobile, Weather, Local Search/Directions, Flight Information, Time, Stock, Date, Travel Suggestion, Restaurant Suggestion \& General Knowledge. This sample list of query examples is shown in the appendix section \ref{query_list}. This task list was provided as another reference for understanding some of the agent's capabilities and the domains supported. \textbf{Participants were NOT required to execute the same tasks specified on the example list.} Participants were free to interact with the prototypes as much as they pleased and execute any task of their choosing.

\subsubsection{Phase 1}
In this phase, the user interacts with \systemname{}  \textbf{at least once} for each of the 10 domains. E.g.:

\begin{dialogue}
    \speak{Researcher} Please ask a query to the system
    
    \speak{Participant} (Asks query)
    
    \speak{System} (Plays response)
    
    \speak{Participant} (Indicates whether the system answered their query correctly)
\end{dialogue}

\subsubsection{Phase 2}
In this phase, the user interacts with Agent Select \textbf{at least once} for each of the 10 domains. E.g.:

\begin{dialogue}
    \speak{Researcher} Please ask a query to the system
    
    \speak{Participant} (Selects desired conversational agent)
    
    \speak{Participant} (Asks query)
    
    \speak{System} (Plays response)
    
    \speak{Participant} (Indicates whether the system answered their query correctly)
\end{dialogue}

In both phases, at least one task is executed for each domain. For each query asked by the user we record the selected agent, asked utterance, agent response, whether the agent responded correctly, and processing latency. Note: \textbf{Some tasks may comprise multiple conversational turns}.

\begin{figure}
\centering
  \includegraphics[width=1\columnwidth]{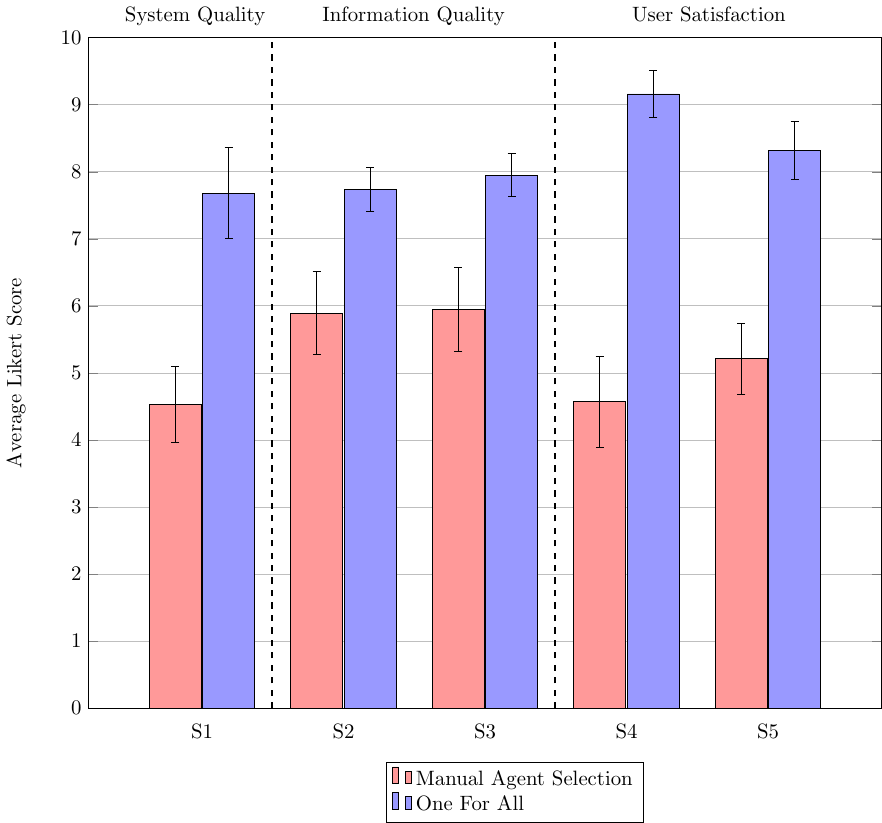}
  \caption{The average scores of the participants' feedback across statements in the questionnaire (higher is better).  Each Statement S1--S5 corresponds to those listed in Table~\ref{tab:questable}.  All results are significant using the Wilcoxon Signed Rank test ($p<.01$).\label{fig:questionnaire}}
\end{figure}

\subsection{System Usability Scale (SUS)}
 The System Usability Scale~\cite{Brooke96sus:a} is a tool for measuring the usability of a system. It consists of a 10-item questionnaire with five response options for respondents; from Strongly agree to Strongly disagree. We utilize the SUS scale \cite{sus-a} to record and calculate participants' individual usability scores.
 
 \subsection{Questionnaire}
 After each phase, participants filled out a questionnaire based on the principles of the DeLone and McLean information-system (IS) success model \cite{delone2003delone, delone1992information}. Each questionnaire consisted of 5 questions designed to evaluate three main aspects of each prototype: System Quality, Information Quality, and User Satisfaction. Questions were slightly adjusted based on the prototype being evaluated (\systemname{} or Agent Select). The evaluation is scoped to evaluate the interaction experience of single-agent use (one agent for all) vs multi-agent (user choice of agents). The participants answered each question on a 10-point Likert scale ranging from 1 = ``strongly disagree'' to 10 = ``strongly agree''. For system quality, participants were asked to evaluate their preference of who should be responsible for agent selection. For example, \textit{``Being able to explicitly select which assistant is useful''}. For information quality, we focused on evaluating if the interactions with the prototype were accurate and usable. For example, \textit{``The assistant is capable of understanding my question and providing a feasible answer''}. For user satisfaction, participants were asked to rate their satisfaction with the prototype and indicate if the experience was to their liking. For example, \textit{``Having multiple assistants make the experience more pleasurable.''} The targeted assessments of each question in the questionnaire are shown in Table \ref{tab:questable}. At the end of each questionnaire, any additional comments, recommendations, or feedback from the participants were also recorded.

\section{Results: User Experience}
In this section, we detail the results of our user study providing insight into users’ perceptions of the interactional experience of each prototype and the prototype's effectiveness in facilitating the completion of user tasks. All reported results in this section are statistically significant using the Wilcoxon Signed Rank Test $(p < .01)$.

Table \ref{tab:sus-accuracy} reports the results of each prototype evaluation. \systemname{} achieved an average SUS score of 86 (SD=8.9) compared to an average of 56 (SD=21.7) for Agent Select. In addition, \systemname{} achieved an accuracy score of 71\% across all tasks executed by participants compared to 57\% by Agent Select. 

\begin{table}
  \centering
  \begin{tabular}{l r r}\toprule
    {\textit{System Prototype}}
    & {\textit{Mean SUS Score ($\sigma$)}}
      & {\textit{Accuracy}} \\
    \midrule
    \systemname{} & 86 (08.9)  & 71\% \\
    Agent Select & 56 (21.7) & 57\%\\
    \bottomrule
  \end{tabular}
  \vspace{0.1cm}
  \caption{User study results showing average system usability score and overall accuracy/correctness of each system prototype.}~\label{tab:sus-accuracy}
\end{table}

\subsection{System Quality}
System quality was evaluated by statement 1 in table \ref{tab:questable}. This statement focused on assessing system acceptability, intelligence, and helpfulness. As shown in Figure \ref{fig:questionnaire}, participants were more receptive to handing off agent selection compared to having the flexibility of choice. \systemname{} achieved a $\mu$ score = 7.68 compared to a $\mu$ = 4.5 for agent select. Participants expressed open-ended feedback on system quality such as: \textit{``Having to pick an assistant is hard because I don't know what each assistant is capable of.``} and \textit{``It would be nice if the best assistant was selected for me.``} 

\subsection{Information Quality}
Information quality was evaluated by statements 2 and 3 in table \ref{tab:questable}. These statements focused on evaluating participants' perception of each prototype's capability of understanding their queries and its ability to provide a feasible and useful response. \systemname{} achieved an average score of $\mu$=7.74 and $\mu$=7.95 on statements 2 and 3 respectively compared to scores of $\mu$=5.89 and $\mu$=5.95 for Agent Select. This score coincides with the reported system accuracy reported in \ref{tab:sus-accuracy} showing that users were more likely to achieve a feasible response to their query when using \systemname{}.

\subsection{User Satisfaction}
Lastly, user satisfaction was evaluated by statements 4 and 5. These statements focused on ascertaining overall user satisfaction and determining each prototype's ability to meet their expectations. Results show that participants experienced greater overall satisfaction using \systemname{}'s interactional experience (S4 $\mu$=9.16) and found it capable of meeting their expectations (S5 $\mu$ = 8.32), as shown in Figure~\ref{fig:questionnaire}.



\section{Quantitative Study Design}
In addition to evaluating the interactional experience of each mode, we further investigate the performance of \systemname{} on completing users' tasks. Given that, as shown in the previous section, users have a significant preference for interaction with a single agent and favor handing off agent orchestration, we investigate and identify a series of key constraints and challenges to designing and employing this mode of orchestration. Our evaluation methodology is further described in the following subsections.

\subsection{Establishing Human Performance}
A large proponent of each prototype's experience is its ability to allow users to execute their tasks. As such, in order to ascertain whether \systemname{} is able to correctly and consistently resolve user tasks, we evaluate its performance against human judgment. We utilize human judgment as the ``\textit{gold standard}'' we want to achieve. From this evaluation, we extract two key metrics: \textit{Accuracy} \& \textit{Response Quality}. Accuracy denoting the percentage of time \systemname{} was able to select the correct agent as denoted by human judges and Response Quality denotes the quality of the selected agent response rated by our human crowd.

\subsubsection{Dataset Collection}
Using Amazon Mechanical Turk and scenario/paraphrasing-based prompts~\cite{Kang:2018}, we crowdsourced utterances across a range of task domains to evaluate \systemname{}. Each utterance denotes a request from the user asking the agent to complete a task in that domain. Domains were selected by observing the capabilities of each of the integrated agents. Our dataset comprises 9 broad domain categories of \textit{Weather, Flight Information, Directions, Weather, Automobile, etc}. Additionally, domains such as automobile contain a range of subdomains such as \textit{fuel economy, adaptive cruise control, etc.}. Workers were paid \$0.12 for 5 utterances. Worker average task completion was ~30 seconds. This rate was commensurate with the U.S. minimum wage at the time this experiment was conducted. After collection utterances were vetted by hand to ensure quality.

\subsubsection{Response Selection Accuracy}
As shown in Figure \ref{fig:mturk} we launched crowdsourcing tasks asking workers to indicate which agent resolved a user's task. We assigned five workers to each selection task and used majority voting to represent the consensus ground truth. In order to prevent selection bias, workers were unaware of which agents provided the response. In addition, the order of options was randomized. Workers were only able to select \textbf{one} of the provided options. This decision was made to have our human judgment process identical to the decision process of \systemname{}. 

In total, we collected 320 evaluation tasks from crowd worker utterances -- 160 from the automobile domain spread across subdomains of (fuel level, adaptive cruise control, lane keeping system, etc), and 160 were evenly spread across the other 8 domains (weather, flight info, etc.). This skew in the domain distribution is due to one of our focuses being deploying \systemname{} in the driving scenario. In particular, we were interested in observing if the other agents affected \systemname{}'s ability to complete tasks when one of the agents is specifically domain focused. For example, if a user tries to complete a task in the driving domain, in a majority of cases, Adasa is the most suited agent to resolve this task. However, a "general purpose" agent such as Google may attempt to resolve this task. We aimed to test \systemname{}'s ability to deal with these scenarios as well as observe crowd workers' ability to distinguish these occurrences as well.

\subsubsection{Response Quality}
\begin{figure}
  \centering
  \includegraphics[width=1\columnwidth]{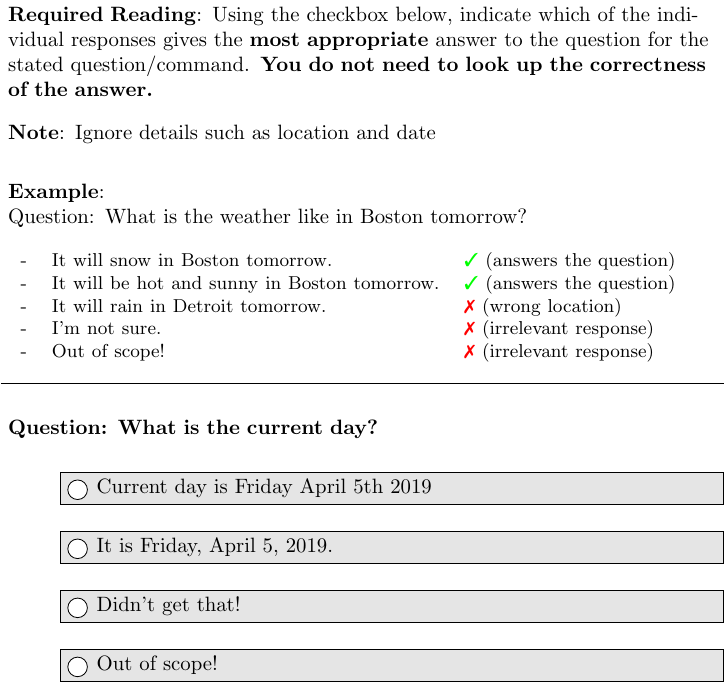}
  \caption{An MTurk task assignment example. 
  We asked workers to decide which of the candidate responses was the most appropriate for the question/command stated.
  This setup allowed us to gather human judgments of the most appropriate responses to inquiries, and also to gather how effective our approach is at deciding on the best responses.
  }~\label{fig:mturk}
\end{figure}

While our human judgment dataset represented the \textit{``best''} outcome for a user task, since workers were only able to select one option as the best it ignored cases where multiple agents were able to also complete the task. Figure \ref{fig:mturk} shows an example of this where the first two agent responses clearly indicate their ability to complete the task. As such, using our crowdsourced dataset of evaluation tasks, we launched around series of crowdsourcing tasks asking workers to rank each agent response using a Likert scale of 1--5 (1 = Very Poor, 2 = Poor, 3 = Acceptable, 4 = Good, 5 = Very Good). This allowed us to not only evaluate the agent's ability to complete the task but assess its ability to clearly indicate its understanding of the user's query and return a clear concise response. 3 workers were assigned per response quality ranking task.

\section{Results: System Performance}
Using the aforementioned dataset and evaluation metrics we evaluated \systemname{} ability to orchestrate multiple agents to complete user tasks. Emphasis is placed on these metrics as they are commonly used to evaluate conversational agents ~\cite{Lin2018Adasa, Hauswald:2015:SOE:2775054.2694347, Fast_2018, sirinumber, ianumber}.

\begin{table*}[ht!]
    \centering
    \begin{tabular}{l|ccc|cccc} \toprule
    & \multicolumn{3}{c|}{Our approaches} & \\
    {Task Domains} & {OFA$_{USE}$} & OFA$_{SIF+GLOVE}$ & OFA$_{BERT}$ & {ALEXA} & {GOOGLE} & {HOUNDFIY} & {ADASA} \\ \midrule
    All & \textbf{70\%} & 60\% & 64\% & 11\% & 21\% & 18\% & 50\% \\
    Automobile & 78\% & 72\% & 76\% & 1\% & 1\% & 0\% & \textbf{99\%} \\
    Weather & 60\% & \textbf{65\%} & 60\% & 15\% & 30\% & 55\% & 0\%  \\
    Time & 55\% & \textbf{60\%} & 50\% & 30\% & 40\% & 30\% & 0\% \\
    Directions & \textbf{50\%} & 50\% & 45\% & 10\% & 45\% & 35\% & 10\% \\ 
    Stock & \textbf{55\%} & 35\% & 45\% & 35\% & 40\% & 25\% & 0\% \\
    Date & \textbf{55\%} & 35\% & 40\% & 35\% & 35\% & 30\% & 0\%  \\
    Travel & \textbf{85\%} & 45\% & 65\% & 35\% & 60\% & 0\% & 5\%   \\
    Flight Info & \textbf{85\%} & 70\% & 70\%  & 5\% & 45\%  & 50\%  & 0\%     \\
    Restaurant & 55\% & 40\% & 45\%  & 10\% & 30\% & \textbf{60\%}  & 0\% \\ \bottomrule
\end{tabular}
\vspace{0.1cm}
    \caption{Breakdown of various Agents vs. Human judgment.  OFA stands for \systemname{}. \textbf{Note:} These percentages indicate the portion of agent responses that were the same as the ground truth human selection. It is possible for an agent to produce a feasible response that is not voted the best by human judgment.}
    \label{tab:humanvsagent}
\end{table*}

\subsection{One For All vs. Human judgment}
As described in the experiment design, we use human judgment to establish our \textit{``gold standard''} dataset. We compare the different \systemname{} response selection engines as well as each of the agents in isolation on our ground truth. Overall \systemname{} across all task domains, \systemname{} using the USE~\cite{cer-etal-2018-universal} response selection engine performed the best by selecting the same agent as the human ground truth \textbf{70\%} of the time. We observe that outside of the restaurant domain, \systemname{} performs better than any single agent isolation. This highlights the advantage of combining distinct agents in multi-agent conversational AI. A detailed breakdown of the results by task domain is shown in Table~\ref{tab:humanvsagent}.

\subsection{Response Quality}
\begin{figure}
\centering
  \includegraphics[width=1\columnwidth]{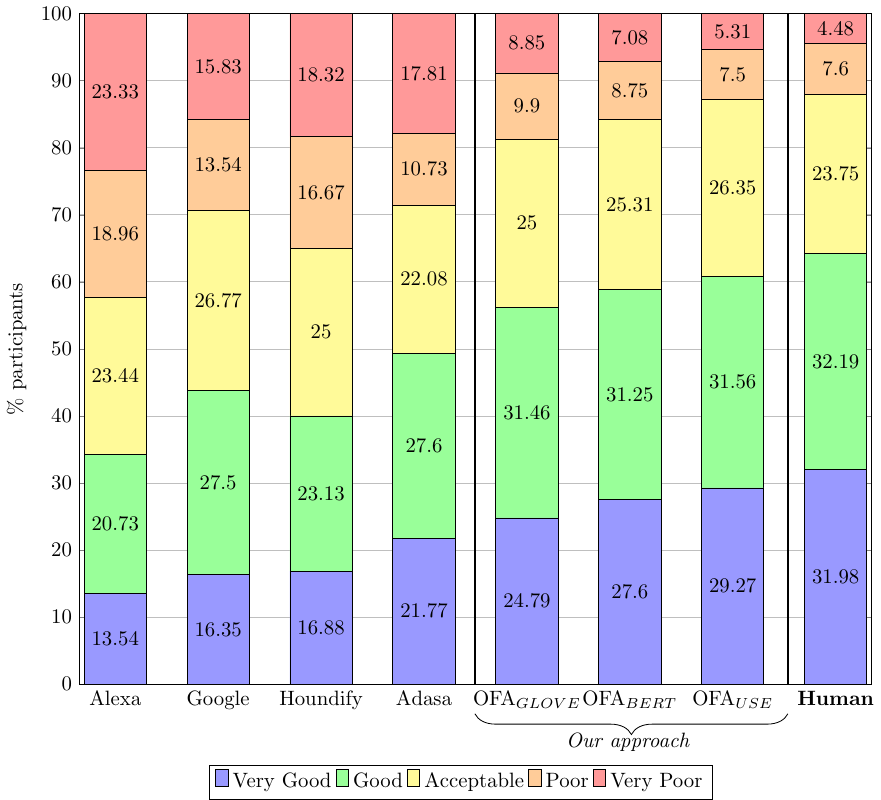}
  \caption{The distribution of agent response quality across Likert scale data points. \systemname{} outperforms each of the assistants in isolation when producing desirable responses and outperforms all assistants in producing the least amount of responses deemed as completely wrong by the crowd. In addition when compared to the human judgment ground truth \systemname{} is par in selecting the most appropriate agent.\label{fig:response-quality}}
\end{figure}

Since \systemname{}'s ranking engine determines the best agent by evaluating the query's semantic relation to the array of agent responses, we sought to evaluate its performance in a similar manner. Expanding on other evaluations which determine the best choice by doing a binary (Y/N) of if the agent completed the task correctly or not \cite{Lin2018Adasa, ianumber, sirinumber}, we take into consideration the other agents did not win the majority vote by measuring the subjective quality of the agent response. We utilized Amazon Mechanical Turk \cite{buhrmester2011amazon} and presented crowd workers with the task as of assessing the quality of each agent response to a given query. For each query three (3) Mturk workers are tasked with ranking the agent response on a 5-point Likert scale (1 = Very Poor, 2 = Poor, 3 = Acceptable, 4 = Good, 5 = Very Good). We launched the response ranking crowdsourcing jobs on a randomized subset of our collected data with a count of 20 jobs per supported domain. 
In total, we launched 320 crowdsourced tasks with three assignments per task allowing us to get a more diverse set of results per domain. From this evaluation we extract the distribution of the response quality of each of \systemname{}'s ranking engines, the individual agents in isolation as well as the human judgment ground truth. Results show that \systemname{} outperformed all of the conversational agents in isolation in completing user tasks and providing appropriate responses. \textbf{87.19\%} of \systemname{}'s agent selections achieved a score of acceptable or better compared to 87.92\% on the human judgment ground truth, showing that \systemname{} is on par with humans at selecting the best agent for a task when privy to each agent's end response. A detailed breakdown is shown in Figure \ref{fig:response-quality}

\section{Discussion}
As indicated by the results. We show that users have a significant preference for using a single-agent interface. Through systematic evaluation, we demonstrate that the single agent interface of \systemname{} is able to resolve users' tasks and provide quality responses that are rated within 1\% of human-selected agent responses. In this section, we discuss a series of key insights \& challenges to consider in the design of multi-agent experiences.

\subsection{Avoiding Non-desirable Agent Responses}
Perception of the quality of an agent's response may differ from person to person. One factor may be the verbiage used or even the length of the agent response \cite{Lin2018Adasa}. However, from our gathered response quality ratings we observe a similarity among the responses denoted as ``poor''. A significant portion of poor responses are responses in which the agent indicated an inability to complete the task. This shows that a key component to facilitating multi-agent use is the avoidance of returning these types of responses, especially in cases where a correct agent response exists. 

In order to evaluate \systemname{}'s ability to avoid these types of responses, we enumerated all the ways in which the agents gave non-desirable responses and highlighted the occurrences in which \systemname{} selected these responses. From observing the collected agent responses we extracted a consistent pattern in the way each of the integrated agents indicates their inability to handle a query. For e.g. Alexa would respond by saying \textit{``I'm not sure''} or \textit{``I don't have an opinion on that''} while Google would provide a more apologetic response like \textit{``sorry I'm not sure how to help'} or \textit{``my apologies I don't understand''}.

As shown in table \ref{tab:bad-responses}, we observe that \systemname{} was able to avoid these responses, reporting a low 3\% rate of occurrence despite the high prevalence of these as a selection option. The high occurrence rate produced by the individual agents is expected due to the broad set of domains evaluated that no single agent could handle in its entirety. The contrast in the percentage of non-desirable responses between the agents with higher rates like Google \& Houndify compared to Alexa \& Adasa shows that they prefer to indicate their inability to resolve your query rather than presenting a response in which they have low confidence. Further inspecting these instances of \systemname{} selecting a ``non-desirable response'', we found that the majority of these were simply a result of \textbf{all} agents returning non-desirable responses. However, we noted an exception in 7 very similar occurrences. For example:

\begin{dialogue}
    \speak{Query} ``Can you explain LKA?''
    
    \speak{Alexa} ``I don't know that one''
    
    \speak{Google} ``Sorry I'm not sure how to help''
    
    \speak{\textbf{Houndify}} \textbf{``Didn't get that''}
    
    \speak {Adasa} ``The Lane Keeping System can help you bring the vehicle back into the traveling lane when your vehicle......''
\end{dialogue}

In the example shown above \systemname{} selected Houndify's response even though Adasa provided the correct response. We find that instances in which the query uses unfamiliar abbreviations and non-standard terms such as ``LKA'', ``ACC'' and ``aid mode'' that a given language model may not represent can present a challenge. In addition, we observe that the difference in query \& response length plays a factor in ranking. These examples show cases where a short and concise question is met with a long detailed response containing lots of extra information. Since \systemname{}'s ranking engine takes into consideration the entirety of the question, this extra information can influence the overall ranking score. As such, we recommend incorporating a pre-filtering method to remove this class of responses from being ranked. Another option would be fine-tuning of the ranking engine model to better recognize these responses and rate them poorly.

\subsection{Agent Flexibility and Agnosticism}
A critical component to facilitating multi-agent interaction is the ability to seamlessly add and remove agents without diminishing the experience or requiring the retraining of any models. Conversational agents may be upgraded to offer a more diverse feature set such as new intent support or improved responses. To facilitate this, we recommend treating each agent as a modular black box. By eliminating the need for any agent specific knowledge you achieve high flexibility, allowing you to \textit{``plug-and-play''} agents as needed while facilitating their individual growth over time at no additional computing cost. Since \systemname{} only considers agents responses for its selection. This allows it to quickly to adapt to any number of agents and their improvements over time.

\subsection{Domain Coverage \& Overlap}
Most commercially deployed conversational agents are engineered to handle domain-specific queries. As such, when acting in isolation their range of support is limited. Seamlessly adding more and more agents allows you to fill in the gaps of non-supported functionality for the other agents in addition to proving fallback when one agent fails. For intelligent personal digital assistants (IPAs) such as Google Assistant and Amazon Alexa which cover a relatively broad spectrum of domains, many of which overlap between agents (e.g., both Google Assistant and Alexa support weather queries), this presents the challenge of determining which agent can be considered as the \textit{most appropriate.} for a given task. When dealing with domain overlap we recommend taking a response selection approach as utilized by \systemname{}. This removes the complexity of designing complicated agent domain mappings that can rapidly change over time. If some form of choice is needed by the user, a simple domain preference list that routes specific tasks to a single agent of the user's choice can suffice.

\begin{table}
  \centering
  \begin{tabular}{l c}
  \toprule
    {\textit{Agent}}
    & {\% Non-desirable Responses} \\
    \midrule
    Alexa & 29.69\% \\
    Google & 48.13\% \\
    Houndify & 56.25\% \\
    Adasa & 36.56\% \\
    \systemname{} & \textbf{3.75}\% \\
    Human & \textbf{2.19}\% \\
    \bottomrule
  \end{tabular}
  \vspace{0.1cm}
  \caption{Breakdown of Non-desirable results by agent. \systemname{} is able to avoid providing these responses to the user.} ~\label{tab:bad-responses}
\end{table}

\section{Conclusion and Future Work}
Although commercial conversational agents promote \textit{``ask me anything''} functionality, they are typically built to focus on a single or finite set of expertise. Given that complex tasks often can require more than one expertise, this results in the users needing to learn and adopt multiple agents to accomplish tasks. One approach to alleviate this is the use of a single-agent interface that abstracts the orchestration of multiple agents in the background unbeknownst to the user. However, this removes the option of choice and flexibility from the user, potentially harming the ability to complete the task. In this paper, we explore the interaction experience of single-agent use (one agent for all) vs multi-agent (user choice of agents) in the context of conversational AI. We design prototypes for each of these use cases, systematically evaluating their ability to provide correct user responses. Using these prototypes, we conduct a series of user studies in which participants execute a series of tasks. Our findings show that users have a significant preference for using a single-agent interface over a multi-agent regarding system usability and system performance. Through systematic evaluation, we demonstrate that multi-agent use is able to provide quality responses that are rated within 1\% of human-selected answers.

Released with our work are two datasets: (1) Our crowdsourced dataset queries containing the agent responses and human ground truth selection. (2) Our user study dataset containing questions asked by participants (to \systemname{} and Agent Select), the returned agent response, and the subjective user satisfaction (yes/no) on the acceptability of the agent response.

We hope that our prototypes and the provision of these datasets will encourage discussion and exciting research on exploring multi-agent usage for simplifying human-agent interaction. 

\section*{Acknowledgements}
We thank our anonymous reviewers for their feedback and suggestions. This work was sponsored by Ford Motor Company.

\bibliographystyle{IEEEtran}
\bibliography{sample-base}

\clearpage
\section{Biography Section}
\vspace{-2cm}
\begin{IEEEbiography}[{\includegraphics[width=1in,height=1.25in,clip,keepaspectratio]{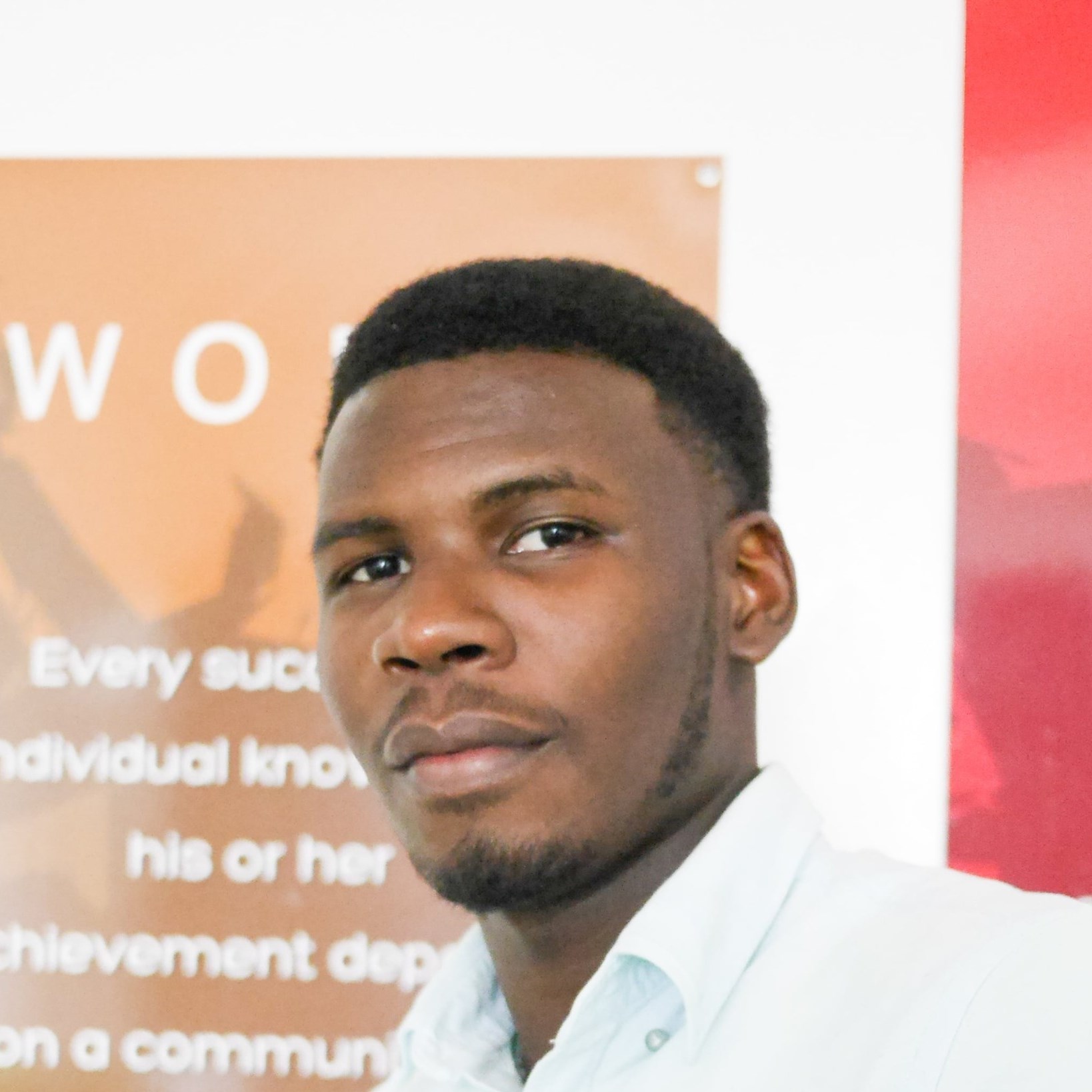}}]{Christopher Clarke}
is a PhD Candidate in Clarity Lab at the University of Michigan. His research is focused on the design of novel systems and approaches optimized for scalable artificial intelligence (AI) infrastructures primarily in NLP \& HAI. Outside of UM, he is a Lecturer I in the Department of Computer Science at the University of Guyana, as well as a Director on the board of Nexus Hub and V75 Inc.
\end{IEEEbiography}

\vskip -2\baselineskip plus -1fil

\begin{IEEEbiography}[{\includegraphics[width=1in,height=1.25in,clip,keepaspectratio]{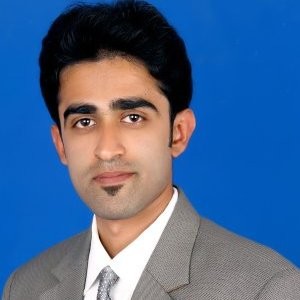}}]{Karthik Krishnamurthy}
is an innovator, and a practitioner in the field of artificial intelligence (AI) and machine learning. He has a Master’s degree in Computer Science and is an active contributor to the industry. As a Machine Learning Scientist at Ford, he takes pride in research and development in areas of Conversational AI, NLP, Computer Vision, and Predictive Intelligence while tackling complex technical problems, and creating solutions that have a positive impact on society.
\end{IEEEbiography}

\vskip -2\baselineskip plus -1fil

\begin{IEEEbiography}[{\includegraphics[width=1in,height=1.25in,clip,keepaspectratio]{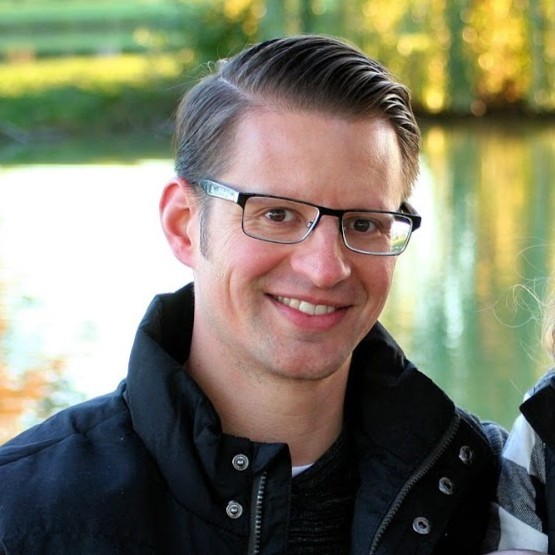}}]{Walter J. Talamonti, Jr.}
is a Research Engineer and Technical Expert at Ford Motor Company Dearborn. He is passionate about translating human performance measures, whether it be in a shared control automated driving setting or one through conversation with an AI or both, into machine learnable actions. His research interests are Affective computing, Artificial Intelligence, Conversational AI, Human Factors, and Advanced Driver Assistance Systems.
\end{IEEEbiography}

\vskip -2\baselineskip plus -1fil

\begin{IEEEbiography}[{\includegraphics[width=1in,height=1.25in,clip,keepaspectratio]{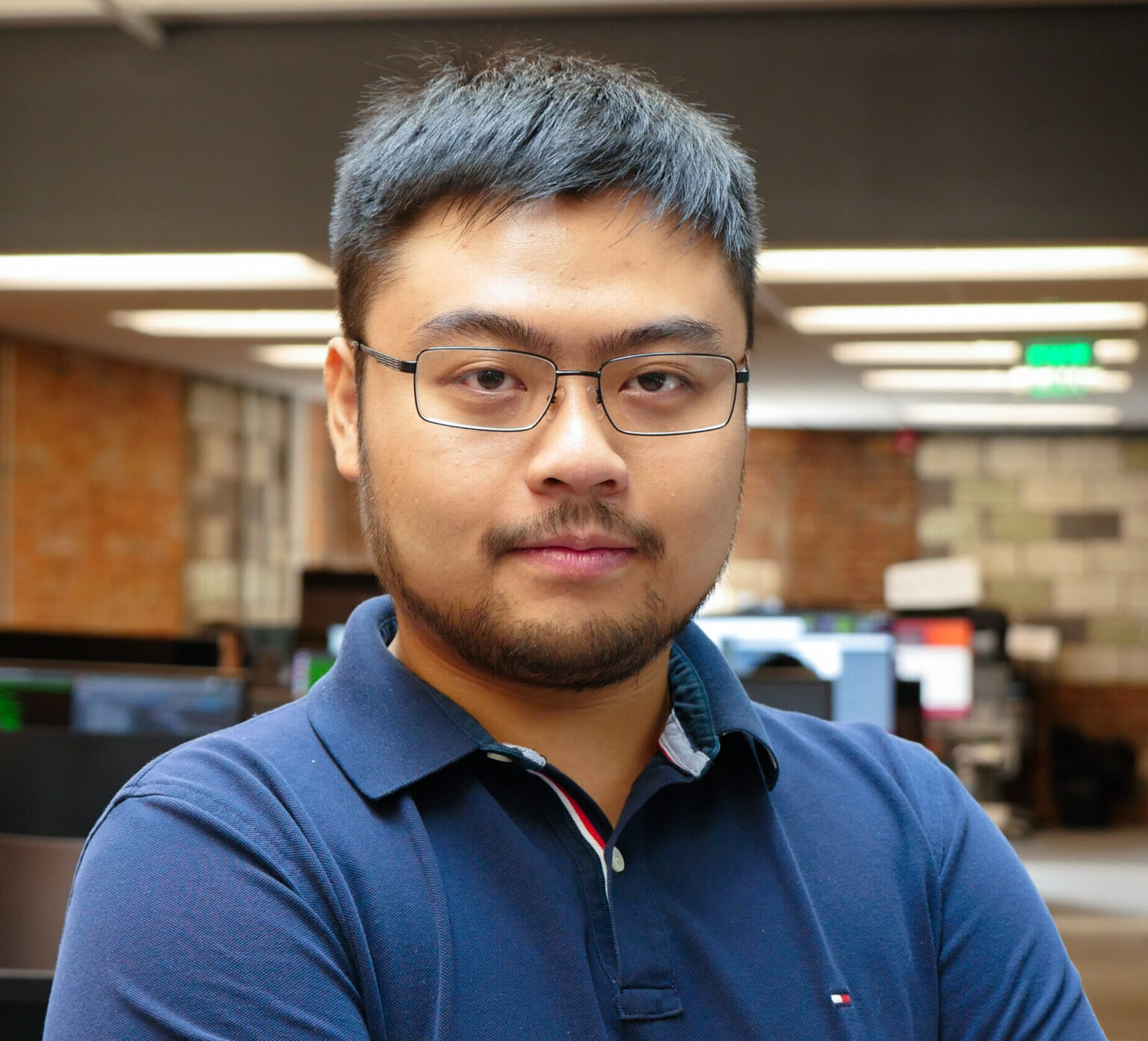}}]{Yiping Kang} is a post-doctoral research fellow in the Computer Science Engineering department at the University of Michigan. He is a research expert in natural language processing, computer vision, computer architecture, and systems, with numerous publications at top-tier academic conferences including ISCA, ASPLOS, ACL, and NeurIPS. He earned his Ph.D. from the University of Michigan in 2018.
\end{IEEEbiography}

\vskip -2\baselineskip plus -1fil

\begin{IEEEbiography}[{\includegraphics[width=1in,height=1.25in,clip,keepaspectratio]{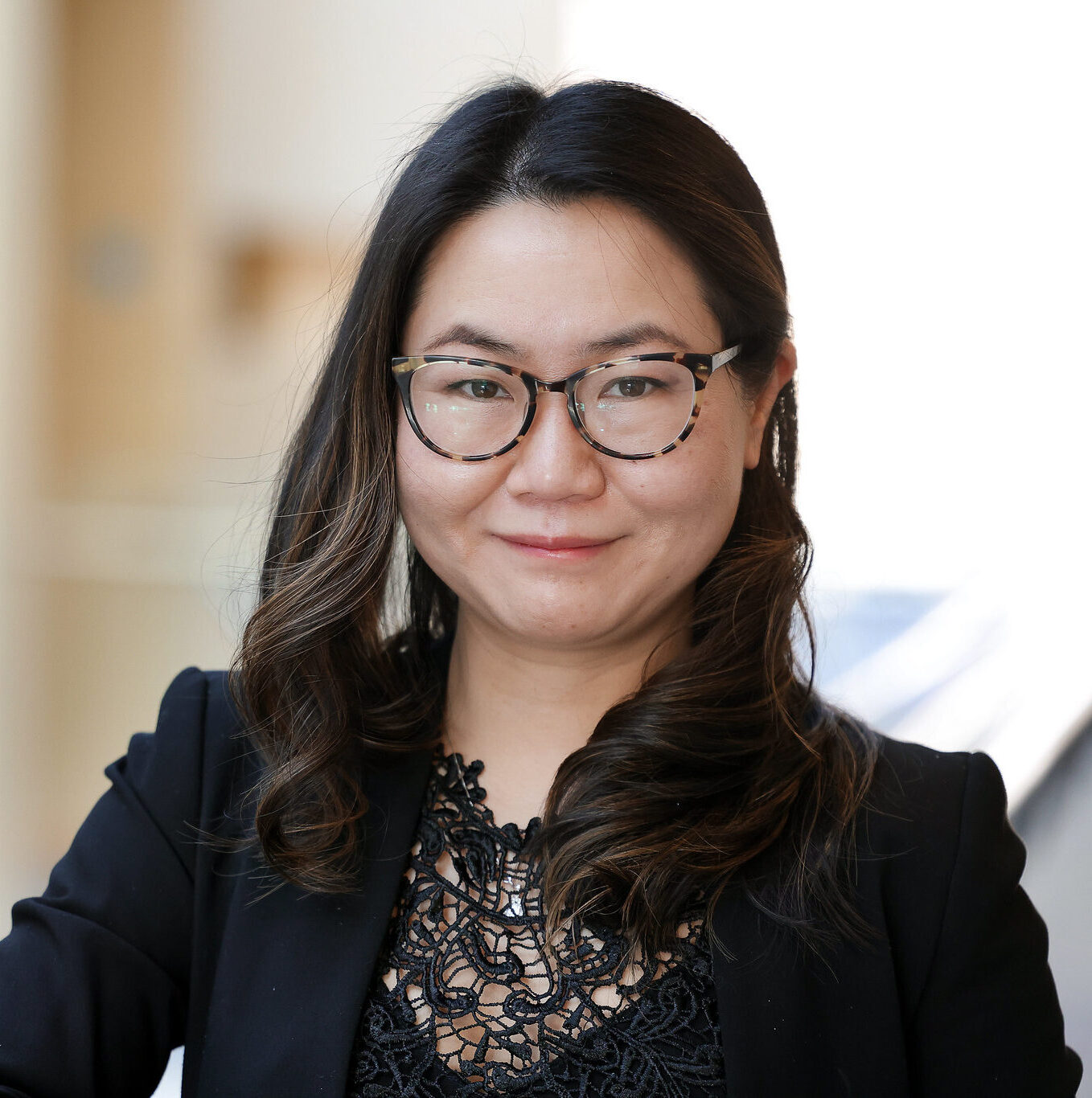}}]{Lingjia Tang} is a professor in the Department of Computer Science and Engineering at the University of Michigan. Her research focuses on designing systems for AI applications. She has received the Google research award, Facebook research award, ISCA Hall of Fame, and MICRO Hall of Fame. She has graduated 10 Ph.D. students and mentored 6 post-doc fellows. Prior to joining UofM, Lingjia received her Ph.D. from the University of Virginia.
\end{IEEEbiography}

\vskip -2\baselineskip plus -1fil

\begin{IEEEbiography}[{\includegraphics[width=1in,height=1.25in,clip,keepaspectratio]{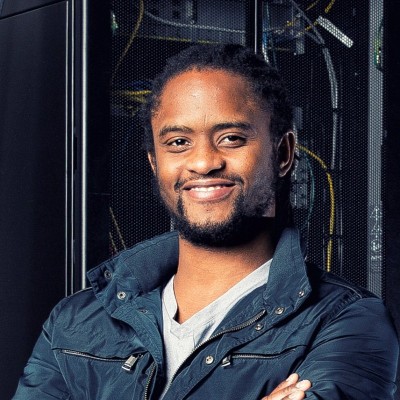}}]{Jason Mars} is a professor of computer science at the University of Michigan, author of the bestselling “Breaking Bots: Inventing A New Voice In The AI Revolution,” and founder of a number of successful AI companies. His work is at the intersection of science, technology, and entrepreneurship with the mission to have a meaningful impact on the lives of every human. 
\end{IEEEbiography}

\clearpage
\appendix

\section{Conversation Agents} \label{agents}

\subsection{Amazon Alexa}
Amazon Alexa~\cite{avs} is Amazon's intelligent voice recognition and natural language understanding service that allows you to voice-enable any connected device that has a microphone and speaker. Using the Alexa Voice Service \cite{avs} HTTP/2 service endpoints we implemented a client package capable of sending and receiving audio interactions with a registered and configured AVS product. Due to the API's audio-only restriction we also implement a Text-to-Speech and Speech-to-Text pipeline into our client package.

\subsection{Google Assistant}
Google Assistant~\cite{googleassistant} is Google's own intelligent personal assistant built with the capability to provides users with a variety of service such as weather, recipes, calendar, booking etc. by understanding natural language utterances spoken by the user. Using the Google Assistant SDK which consists of the Google Assistant Library \cite{googleassitantlib} and the Google Assistant Service we configured implemented a client package that allows us to interface with Google Assistant's gRPC framework \cite{googleassitantservice} to send queries to Google Assistant. While Google supports both audio and text input/output attached to our package is a Speech-to-Text pipeline as it was observed that Google does not always return a text transcription of their responses.

\subsection{Houndify}
Sound Hound Inc.'s Houndify~\cite{soundhound} is a platform that lets anyone add smart, voice-enabled, conversational interfaces to anything with an internet connection. Houndify comes prepackaged with what are called \textit{Houndified Domains}. These domains are programs that allow the Houndify server to respond to queries for a certain topic. We deployed and configured our own Houndify application with some of Houndify's prepackaged domains such as \textit{weather, local search, wikipedia} etc. Using the Houndify Client SDK \cite{houndifysdk} we implemented a package that lets us interface with our Houndify application.

\subsection{Adasa} 
Based on prior work~\cite{Lin2018Adasa}, we used Clinc's~\cite{clinc} conversational AI platform to rebuild and extend upon Adasa, an in-vehicle conversational assistant for driving features.
Adasa allows the drivers to access and control in-vehicle features such as Lane Keeping System and Adaptive Cruise Control in a conversational manner via voice.
In addition, we extended Adasa to incorporate extra features such as navigation, maintenance appointment scheduling, fuel level, and car indicator information.
For example, a our adaptation of Adasa can handle queries such as \textit{``Can I get to Baltimore with the fuel I have?''} or \textit{``Activate cruise control and set my distance to 2 car units''} or \textit{``Which side of the car is my fuel door on?''}

\section{User Study Materials}

\section{Sample task list} \label{query_list}

\subsection{Automotive}
\begin{enumerate}
    \item Activate the adaptive cruise control, set distance to be one car unit
\item Can my adaptive cruise control alert me if I drift out of my lane?
\item Can I turn off the seatbelt light on my dashboard?
\item Can you tell me which side of the car the fuel door is located on, please?
\item Can I at least get to Baltimore with the gas I have
\end{enumerate}

\subsection{Weather}
\begin{enumerate}
\item what's the weekly weather report for the city
\item Is it humid in Compton now?
\item What will the highest temperature in San Jose be in the next 7 days?
\item will it snow tomorrow
\item show me the weather forecast for the week
\end{enumerate}

\subsection{Time}

\begin{enumerate}
\item What time is it in San Jose?
\item When does the sun set today?
\item Please give me the time in tanzania at this moment
\item I need you to tell me what time it is in new york nowDid 
\item What time is it in the eastern timezone?
\end{enumerate}

\subsection{Directions}
\begin{enumerate}
\item Find me a chinese spot i could get to the quickest.
\item Are there shopping centers nearby?
\item Are there any hospitals near me
\item I need a place that serves coffee locally.
\item Where is there a pizza place near me?
\end{enumerate}

\subsection{Flight Info}
\begin{enumerate}
\item How much is a one way flight to new york leaving on Friday?
\item What’s the cost of a plane ticket to Brazil?
\item Are there any one way american airline flights to miami leaving friday available
\item is there any one way flight to New York for tomorrow available
\item Cheap flights to the bahamas
\end{enumerate}

\subsection{Stock}
\begin{enumerate}
\item Did the Dow Jones go up or down today?
\item What times does the market open today?
\item What is apple trading at right now?
\item How did microsoft do for today?
\item What was the closing price of apple?
\end{enumerate}

\subsection{Date}
\begin{enumerate}
\item What is the current day?
\item What is the date in 5 days?
\item I need to know tomorrow's date
\item Please tell me what today is
\item Date please
\end{enumerate}

\subsection{Travel}
\begin{enumerate}
\item Is there anything fun to do in berlin?
\item What fun is there to do in england?
\item What are the most popular attractions in gatlinburg?
\item What are some things I can do in portland?
\item Could you tell me what fun tourist things I could do in tokyo?
\end{enumerate}

\subsection{Restaurant}
\begin{enumerate}
\item What are the best restaurants open tonight?
\item I need some suggestions for dinner places tonight
\item Can you suggest a thai restaurant, please?
\item I need reviews for places serving tacos in chicago
\item In Cleveland, are there any good places that serve clams?
\end{enumerate}

\subsection{General Knowledge}
\begin{enumerate}
\item Which college did Barack Obama go to?
\item When do the NBA finals begin?
\item Who is the president of the United states?
\item When is the next Lakers game?
\item How many miles are in a 5k?
\end{enumerate}

\end{document}